\documentstyle[psfig2]{mn}

\newif\ifAMStwofonts
\ifoldfss
  \ifCUPmtlplainloaded \else
    \NewTextAlphabet{textbfit} {cmbxti10} {}
    \NewTextAlphabet{textbfss} {cmssbx10} {}
    \NewMathAlphabet{mathbfit} {cmbxti10} {} % for math mode
    \NewMathAlphabet{mathbfss} {cmssbx10} {} %  "   "    "
  \fi
  \ifAMStwofonts
    \ifCUPmtlplainloaded \else
      \NewSymbolFont{upmath} {eurm10}
      \NewSymbolFont{AMSa} {msam10}
      \NewMathSymbol{\upi}     {0}{upmath}{19}
      \NewMathSymbol{\umu}     {0}{upmath}{16}
      \NewMathSymbol{\upartial}{0}{upmath}{40}
      \NewMathSymbol{\leqslant}{3}{AMSa}{36}
      \NewMathSymbol{\geqslant}{3}{AMSa}{3E}

    \fi
  \fi
\fi % End of OFSS

\ifnfssone
  \newmathalphabet{\mathit}
  \addtoversion{normal}{\mathit}{cmr}{m}{it}
  \addtoversion{bold}{\mathit}{cmr}{bx}{it}
  \newmathalphabet{\mathbfit} % math mode version of \textbfit{..}
  \addtoversion{normal}{\mathbfit}{cmr}{bx}{it}
  \addtoversion{bold}{\mathbfit}{cmr}{bx}{it}
  \newmathalphabet{\mathbfss} % math mode version of \textbfss{..}
  \addtoversion{normal}{\mathbfss}{cmss}{bx}{n}
  \addtoversion{bold}{\mathbfss}{cmss}{bx}{n}
  \ifAMStwofonts
    \ifCUPmtlplainloaded \else
      \UseAMStwoboldmath
      \makeatletter
      \new@mathgroup\upmath@group
      \define@mathgroup\mv@normal\upmath@group{eur}{m}{n}
      \define@mathgroup\mv@bold\upmath@group{eur}{b}{n}
      \edef\UPM{\hexnumber\upmath@group}
      \new@mathgroup\amsa@group
      \define@mathgroup\mv@normal\amsa@group{msa}{m}{n}
      \define@mathgroup\mv@bold\amsa@group{msa}{m}{n}
      \edef\AMSa{\hexnumber\amsa@group}
      \makeatother
      \mathchardef\upi="0\UPM19
      \mathchardef\umu="0\UPM16
      \mathchardef\upartial="0\UPM40
      \mathchardef\leqslant="3\AMSa36
      \mathchardef\geqslant="3\AMSa3E
    \fi
  \fi
\fi % End of NFSS release 1

\ifnfsstwo
  \DeclareMathAlphabet{\mathbfit}{OT1}{cmr}{bx}{it}
  \SetMathAlphabet\mathbfit{bold}{OT1}{cmr}{bx}{it}
  \DeclareMathAlphabet{\mathbfss}{OT1}{cmss}{bx}{n}
  \SetMathAlphabet\mathbfss{bold}{OT1}{cmss}{bx}{n}
  \ifAMStwofonts
    \ifCUPmtlplainloaded \else
      \DeclareSymbolFont{UPM}{U}{eur}{m}{n}
      \SetSymbolFont{UPM}{bold}{U}{eur}{b}{n}
      \DeclareSymbolFont{AMSa}{U}{msa}{m}{n}
      \DeclareMathSymbol{\upi}{0}{UPM}{"19}
      \DeclareMathSymbol{\umu}{0}{UPM}{"16}
      \DeclareMathSymbol{\upartial}{0}{UPM}{"40}
      \DeclareMathSymbol{\leqslant}{3}{AMSa}{"36}
      \DeclareMathSymbol{\geqslant}{3}{AMSa}{"3E}
    \fi
  \fi
\fi % End of NFSS release 2

\ifCUPmtlplainloaded \else
  \ifAMStwofonts \else % If no AMS fonts
    \def\upi{\pi}
    \def\umu{\mu}
    \def\upartial{\partial}
  \fi
\fi

\title{Searching for Stars in Compact High-Velocity Clouds. II}
\author[Hopp et al.]
       {U. Hopp$^{1, 2}$,\thanks{\sf Based on observations collected 
at the European Southern Observatory, Chile, during run P070.B-0206A.} 
       R.E. Schulte-Ladbeck$^3$,
       J. Kerp$^4$ \\
       $^1$Universit\"ats-Sternwarte M\"unchen, Scheinerstr. 1,
       D-81679 M\"unchen, Germany \\
       $^2$Max--Planck--Institut f\"ur Extraterrestrische Physik,
       Giessenbachstr., D-85748 Garching, Germany \\
       $^3$University of Pittsburgh, Department of Physics \&
       Astronomy, Pittsburgh, PA 15260, USA \\
       $^4$Radioastronomisches Institut der Universit\"at Bonn, Auf
       dem H\"ugel 71, D-53121 Bonn, Germany} 
\date{Accepted .
      Received ;
      in original form }

\pagerange{\pageref{firstpage}--\pageref{lastpage}}
\pubyear{2002}

\begin{document}

\maketitle

\label{firstpage}

\begin{abstract}
We address the hypothesis that High Velocity Clouds correspond to the
``missing" dwarf galaxies of the Local Group predicted by cosmological
simulations. To this end, we present optical and near-infrared
photometry of five additional High Velocity Clouds, one of which
produces Lyman series absorption on the sight line towards the Quasar
Ton S210, with sufficient resolution and sensitivity to enable the
detection of an associated stellar content. We do not detect
significant stellar populations intrinsic to any of the five
clouds. In combination with the results from our paper~I, which had
yielded non detections of stellar content in another five cases, we
find that there is a 50\% chance of getting a null result in ten
trials if fewer than 7\% of all High Velocity Clouds contain stars.
We conclude that the population of High Velocity Clouds is an unlikely
repository for the ``missing" dwarfs of the Local Group.
\end{abstract}


\section{Introduction}

Simulations of structure formation predict the existence of hundreds
of small satellite halos in large parent halos (Klypin et al. 1999,
Moore et al. 1999). Accordingly, and unless low mass satellite halos
are free of gas and stars, a structure such as the Local Group should
contain several hundred dwarf galaxies. The actually observed number
is an order of magnitude smaller (Mateo 1998). This has given rise to
the ``missing" dwarfs problem of cosmological simulations.

The Local Group dwarfs may still be there but may have been overlooked by
previous observations. This is the basic premise of our
work. Specifically, Blitz et al. (1999) proposed that the ``missing"
dwarfs could be associated with High Velocity Clouds (HVC). HVCs are
parcels of gas with extremely high radial velocities that are
inconsistent with Galactic rotation models. Unfortunately, it has
proven very difficult to determine distances for HVCs (e.g., Wakker
\& van Woerden 1997; Braun 2001; Putman 2006). In the Blitz et
al. hypothesis HVCs are extragalactic with a mean characteristic
distance of about 1~Mpc, rather than clouds in the Galactic halo. They
are considered to be leftover building blocks that are in-falling into
the Local Group.
 
 The suggestion that HVCs could be the``missing"
dwarf galaxies of the Local Groups has spurred several searches for
their stellar content.  Searches for stars in HVCs have either looked
for over-densities above Galactic foreground in the stellar
distributions towards HVCs, or, if data of sufficient resolution were
available in two or more filters, they have employed colour-magnitude
diagrams (CMD) in an attempt to measure distances from stellar
population ``standard candles".  
 
 High-surface-brightness galaxy
counterparts to HVCs were quickly ruled out (Braun \& Burton
1999). Simon \& Blitz (2002) inspected digitized Palomar Sky Survey
(POSS) plates for 264 northern HVCs with a spatial filtering
method. The surface brightness limits reached by their analysis (e.g.,
26 magnitudes~arcsec$^{-2}$ in V) would have recovered all known Local
Group galaxies with the exception of four nearby and extended dwarf
Spheroidal galaxies. Similarly, Sloan Digital Sky Survey data of 13
HVCs (including one compact HVC, hereafter CHVC) have not revealed any
stellar over-densities which could be attributed to an intrinsic
stellar content (Willman et al. 2002).

\begin{table*}
 \centering
 \begin{minipage}{140mm}
  \caption{(Compact) High Velocity Clouds observed with VLT/FORS. N
  stars is the number of point sources detected in both, V and I.} 
  \begin{tabular}{lrrrrr}
Identifier &  RA [h:m:s]  & Dec [$^o$:':"] & A$_V$ [m] & A$_K$ [m] & N stars \\

CHVC 039.4-73.6-165  &  23:39:48 & -24:08:00 & 0.061 & 0.007 & 165 \\
CHVC 224.0-83.4-197  &  01:20:42 & -28:12:00 & 0.052 & 0.006 & 198 \\
 HVC 229-74-168      &  02:02:46 & -30:11:00 & 0.066 & 0.007 & 190 \\
CHVC 225.1-42.5+175  &  04:26:00 & -26:31:00 & 0.145 & 0.016 & 234 \\
 HVC 225-42+190      &  04:29:48 & -26:00:00 & 0.131 & 0.014 & 503 \\ 
 
\end{tabular}
\end{minipage}
\end{table*}

Optical CMD searches for stars in HVCs have been described in Braun
(2001), Davies et al. (2002), and Siegel et al. (2005), while a
near-infrared CMD search toward one cloud has been discussed in Simon
et al. (2006). All of these works, further discussed in section 4,
indicate that there is no resolved stellar content in HVCs.
 
Hopp, Schulte-Ladbeck \& Kerp (2003, hereafter paper I) combined deep
optical and archival near-infrared photometry of stellar distributions
toward four CHVCs and one HVC. This observational approach allowed us
to conduct a systematic search for stars a) within a field of view
that was much smaller than the size of the HI distributions but with a
sensitivity that reached twice the characteristic distance of 1~Mpc
proposed by Blitz et al. (1999) and b) within a field of view that
covered the HI extent of the clouds but with a sensitivity that
reached only half of the characteristic distance at best. We showed
that our technique is very sensitive to low-surface-brightness
populations, reaching a V-band surface brightness of 29 magnitudes
arcsec$^{-2}$. No stars associated with the CHVCs or with the HVC were
detected. We concluded that the non detections of stellar content
associated with five (C)HVCs had a 50\% chance of occurring if the
true fraction of (C)HVCs which contains stars is less than 13\%. Here
we present an extension of our work from paper~I to an additional five
objects, and discuss the statistical implications of the results.

\section{Data}
The optical data were gathered in continuation of our service observing
project at the European Southern Observatory's (ESO) Very Large Telescope
(VLT) facility. The data were collected in fall of 2002 and summer of 2003.
Our second observing sample was approved for an additional five targets. The
target list is presented in Table~1. The targets were selected from the
catalogues of Braun \& Burton (1999) and Putman et al. (2002). The objects
from the Putman et al. paper are all classed CHVC in that catalogue. Note that
Braun \& Burton (1999) select as a CHVC a cloud whose integrated HI~emission
at the half power contour has an area of less than four square degrees. It
should be noted that the area of the optical detector used in this work had an
area of only 50 square arc minutes (or 0.01 square degree) while the data
extracted from the Two Micron All Sky Survey encompassed an area of about
three square degrees. The pointings/extractions were centered on the highest
column density regions of the clouds (see below). Given the experience gained
in our previous observing run, we selected targets that have very high
Galactic latitudes, high HI column densities, well resolved radio
observations, and no bright foreground stars in the immediate vicinity. The
pointing positions, listed in Table~1, are from new observations with the
Effelsberg 100-m radio telescope (for the objects labeled HVC, Westmeier 2003)
or were determined using literature HI data. These accurate positions are
critical since the detector used for the optical imaging has a small field of
view, covering only a small fraction of the CHVCs towards their very centers,
along the lines of sight with the highest HI column densities.

We used the FORS~1 instrument which has a field of 6\farcm8 x 6\farcm8. The
observing parameters were the same as in our previous run; as was the
reduction procedure (see paper~I). We again requested seeing
conditions of $<$0\farcs6. This is essential in order to separate
resolved stars from background galaxies at the magnitudes at which we
expect to find stellar populations in the (C)HVCs. The integration
times were 5~$\times$~300s in I and 3~$\times$~300s in V. The limiting
magnitudes, which we define as the magnitudes at which the photometric
errors equal 0.1~mag, were V$\sim$26.25, I$\sim$ 24.50. In generating
the final source catalogues, we again excluded clearly extended sources
with sizes larger than the PSF, presumed to be mainly background
galaxies. The last column in Table~1 (N stars) lists for each (C)HVC,
the number of objects classed to be stars and detected in V and I.
The optical CMDs of the (C)HVCs are qualitatively similar to one
another and to those shown in paper~I. As the new targets are located
at even higher Galactic latitudes than the ones observed before, N
stars is much smaller; and the CMDs are much less
crowded. Specifically, our previous target list had Galactic latitudes
just above $|$30$^o$$|$; three of the targets in Table~1 are located
above $|$70$^o$$|$. Table~1 lists the Galactic absorptions in V and K
for our objects from Schlegel et al. (1998, obtained using
NED\footnote{The NASA/IPAC Extragalactic Database 
 (NED) is operated
by the Jet Propulsion Laboratory, California Institute of
Technology,
 under contract with the National Aeronautics and Space
Administration.}), which is also smaller than the absorptions toward
the targets observed in the first run. In Figure~1, we display the
[V-I,~I] CMD of one of the CHVCs, CHVC 224.0-83.4-197, from the new
sample. There are fewer objects than N stars plotted on the figure,
because we show only those data for which the photometric errors in V,
I are less than 0.1~mag. We do not show the CMDs of the other
(C)HVCs, because their fan-shaped stellar distributions
qualitatively resemble the one displayed in Fig.~1. 

\begin{figure}
\centerline{\psfig{figure=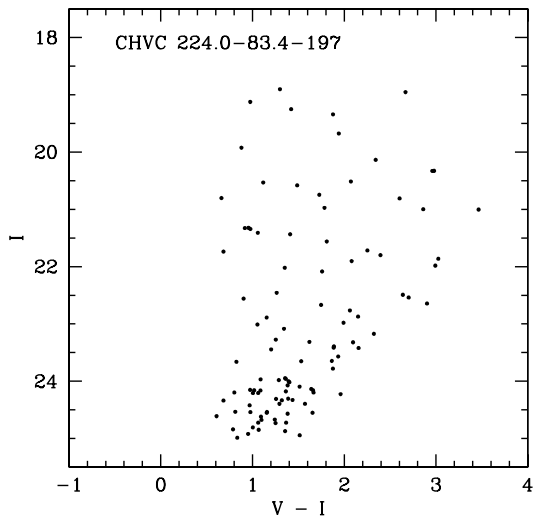,width=8.0cm,angle=0}}
\caption{The V, I CMD of CHVC 224.0-83.4-197. The CMDs of the other 
targets are qualitatively similar.
 \label{figlabel}}
\end{figure}

The Two Micron All Sky Survey\footnote{The 2MASS project is a
collaboration between The University of Massachusetts and the Infrared
Processing and Analysis Center (JPL/Caltech). Funding is provided
primarily by NASA and the NSF.} covers 99.998\% of the sky. The FHWM
of the 2MASS PSF is less than 0.8" in all bands. 2MASS point-source
photometry achieves a 10-$\sigma$ (0.109~mag) detection at J = 15.8, H
= 15.1, and K$_S$ = 14.3 for un-confused sources outside of the
Galactic Plane. Whereas we used the second incremental data release of
the 2MASS point source catalogue in paper~I, here we use the final 2MASS
archive. The 2MASS data were extracted via the Infrared Processing and
Analysis Center (IPAC) WWW site. As in paper~I, we culled point
sources within a 1~degree radius of the positions listed in
Table~1. The 2MASS fields thus cover most of the HI distributions of the
(C)HVCs.

We produced near-infrared position plots and CMDs. The position plots
and CMDs for the new targets differ from those of the previously
studied ones only in that they contain fewer foreground stars due to
higher Galactic latitudes. We provide one sample position plot
(Figure~2), and one sample [J-K$_S$,~K$_S$] CMD (Figure~3), choosing
again CHVC 224.0-83.4-197. At the time we wrote paper~I, there were no
2MASS data available as yet for one of the sample objects,
HIPASS~J1712-64. Data for this HVC were extracted from the final
archive, and inspected along with those of the new targets. The
position plots and near-infrared CMDs of HIPASS~J1712-64 are
qualitatively similar
to those of other (C)HVCs at comparable Galactic latitudes.

\begin{figure} 
\centerline{\psfig{figure=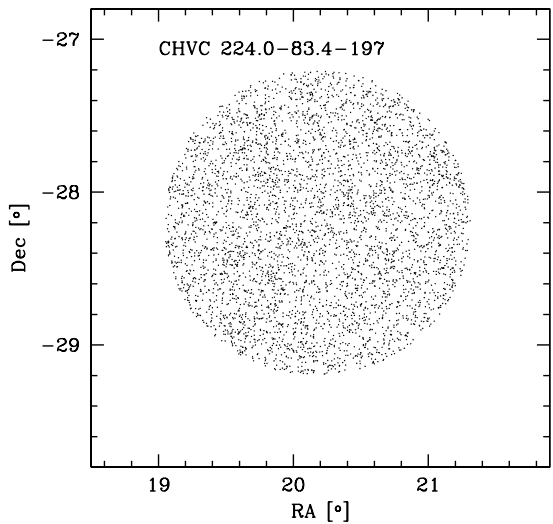,width=8.0cm,angle=0}} 
\caption{A plot of the positions of all sources detected in J and 
K$_S$ toward CHVC 224.0-83.4-197. The position plots of the other 
targets are qualitatively similar.
 \label{figlabel}}
\end{figure}

\begin{figure}
\centerline{\psfig{figure=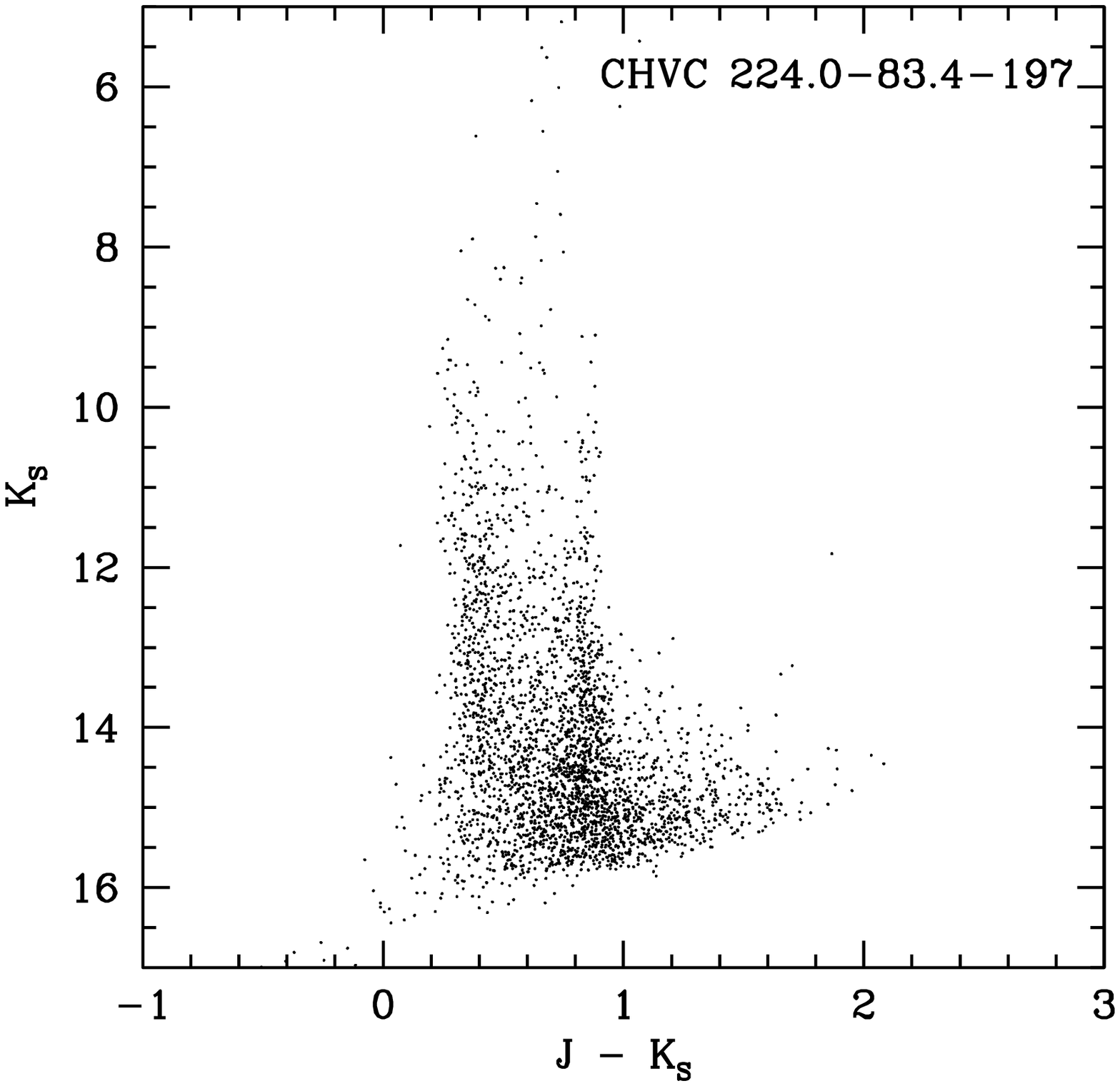,width=8.0cm,angle=0}}
\caption{The J, K$_S$ CMD of CHVC 224.0-83.4-197. The CMDs of 
the other targets are qualitatively similar.
 \label{figlabel}}
\end{figure}

\begin{table*}
 \centering
 \begin{minipage}{80mm}
  \caption{(Compact) High Velocity Clouds which have been searched 
for resolved stars}
  \begin{tabular}{lllr}
Team            & V Limits & K Limits & Identifier       \\
                                \hline
Davies et al. 2002 & 23.3 & N/A & HVC 125+41-207 \\
                    &      & & HVC 070+51-146 \\
                    &      & & HVC 018+47-145 \\
\hline
Paper I             & 25-26& 15.5 & HIPASS~J1712-64  \\
                    &      & & HVC032-31-299    \\
                    &      & & HVC039-31-265    \\
                    &      & & HVC039-33-260    \\
                    &      & & HVC039-37-231    \\ 
\hline
Siegel et al. 2005 & 21.8 & N/A & CHVC 017-25-218  \\ 
                    &      &  & HVC 030-51-1119 \\
                    &      & & CHVC 271+29+181  \\
                    &  21  &  & HVC 267+26+216  \\
\hline
Simon et al. 2006   & N/A & 15 & Complex H\\
\hline
This work           &  26.25  & 15.5 & CHVC 039.4-73.6-165  \\
                    &      & & CHVC 224.0-83.4-197  \\
                    &      &  & HVC 229-74-168      \\
                    &      & & CHVC 225.1-42.5+175  \\
                    &      &  & HVC 225-42+190      \\
\hline
\end{tabular}
\end{minipage}
\end{table*}

\section{Results}

The (C)HVCs we targeted are located at high Galactic latitudes. Therefore, the
contribution by red Galactic foreground stars to the data is small. This has
benefits in particular for distinguishing late-type stars, such as red giant
branch (RGB) stars or asymptotic giant branch (AGB) stars, intrinsic to the
(C)HVCs, from Galactic foreground stars. Indeed, the number of sources
detected in the optical, N stars from Table~1, for example, is factors of two
to five smaller compared to what we saw in paper~I. Given that the data were
obtained with the same instrumentation and under similar conditions, we
attribute this drop in N stars to the fact that the Galactic foreground for
the targets observed here, is factors of two to five smaller than it was for
our first sample. A model of the stellar distribution perpendicular to the
galactic disk follows an exponential drop in stellar density (e.g. Kuijken \&
Gilmore, 1989), which can be well approximated by a plan-parallel description
for the earth-bound observer. Thus, to first order, the surface density of
disk stars should vary as {\it cos($b$)} with $b$ the Galactic latitudes (here
between $\approx$ 30$^o$ and $\approx$ 75$^o$).  This implies variations of a
factor of $\approx$~3.3 in good agreement with our observations. A more
realistic approach is given by the so-called Besan\c{c}on models of Robin et
al. (2003).  We used the web page calculator of these models to verify that
the trend in number counts with galactic latitude shown by these models agree
pretty well with our $I$-band observations.

The [V-I,~I] CMDs of the (C)HVCs in the new sample are qualitatively 
similar to what we observed in paper~I. This can be seen by comparing
Fig.~1 in this paper, to Fig.~1 in Hopp et al. (2002).
There is a fan-shaped distribution of stars
bound by the detection limits and by V-I$\approx$1; and the number of
stars increases toward fainter magnitudes. Most significantly, the
CMDs of the (C)HVCs are quite different from those of known Local
Group dwarf galaxies: we notice neither a young, blue plume, nor an
old red, plume of stars on the CMDs of the (C)HVCs.

A plume of stars with V-I$<$1 indicates the presence of a young stellar
population. Its absence on all of the CMDs is evidence for the absence
of young main sequence (MS) and blue supergiant (BSG) stars in the
(C)HVCs. The brightest evolutionary phase of stars with ages larger
than about one Gyr occurs on the RGB. The RGB stars, if present,
should be visible as a vertical red plume of stars extending over
about 4~mag. in I. The tip of the red giant branch (TRGB) has M$_I$ of
about --4 for stars of a wide range of metallicities. It is a well
calibrated distance indicator (Lee, Freedman \& Madore 1983); and, at
a distance of 1~Mpc, it would appear at I$\approx$21. A TRGB for the
full range of stellar metallicities would have been detected to I
magnitudes of at least 22.5, or to a distance of up to 2~Mpc, in the
(C)HVCs listed in Table~1. Since no RGBs are obvious in any of the
data sets, a straightforward conclusion is that there are no intrinsic
populations of RGB stars in the (C)HVCs, and that the observed sources
are mainly due to stars in the Milky Way and a few remaining
unresolved background galaxies.

Since the Galactic foreground is comprised predominantly of red stars, the
higher galactic latitudes of the new fields were not helpful in increasing our
sensitivity to intrinsic blue stars. In contrast, the sensitivity of our new
data for intrinsic, red stars improved by factors of two to five. In paper~I,
we carried out simulations to investigate how much of a stellar population we
could add to the CMDs before we would notice the presences of blue or red
plumes of stars indicative of young or old stellar populations intrinsic to
the CHVCs. This served as an indication of the sensitivity limit of our
survey.  For the new datasets, we find that any putative young or old stellar
populations contribute between zero and a few percent of stars detected on the
optical CMDs. This indicates a sensitivity limit in V of
30~magnitudes~arcsec$^{-2}$ (cf. paper~1, section 2.3).

The 2MASS position plots for the newly observed (C)HVCs show evenly
distributed point sources across the fields without the kind of central source
density increase which one would expect to see if there were stellar
populations at the locations of the HI maxima of the (C)HVCs (see Fig.~2).
This is despite the fact that the low Galactic foreground contributions favour
the detection of smaller numbers of intrinsic stars.

Two of the fields exhibit one off-center cluster of about a dozen sources
each, with colours J-K$_S$ between 1 and 1.4 and at magnitudes where we might
expect to detect the AGB of a stellar population intrinsic to the (C)HVCs.
Upon further inspection, we find that both features can be explained with
unresolved background galaxies belonging to known galaxy clusters. The CHVC
225.1-42.5+175 field exhibits these objects at RA=67.49, DEC=-26.38 and with
14$<$K$_S$$<$16. The position concides with the galaxy cluster Abell~0495. The
field of CHVC 039.4-73.6-165 shows a clustering of sources at RA=354.25,
DEC=-24.19 with 15$<$K$_S$$<$16. The galaxy cluster Abell~2628 is located at
this position; its member galaxies have redshifts of around 0.185. We checked
that the near-infrared magnitudes and colours of the sources that we detect
with 2MASS can be explained with those of galaxies using evolutionary models
(e.g. Maraston 1998).

The 2MASS CMDs exhibit the same features as those shown in paper~I, displaying
two pronounced plumes of stars at J-K$_S$ of about 0.3 and 0.8, and a less
populated one at colours in between (Fig.~3).  In contrast, dwarf galaxies of
the Local Group show the blue plume of young stars near J-K$_S$$\approx$0, as
well as giant branches that are easily detected via horizontal fingers of AGB
stars with colors $>$1.4. The near-infrared CMDs of the (C)HVCs do not
resemble those of Local Group dwarfs, but rather, that of Galactic stars.

As discussed in paper~I, 2MASS data are sensitive to young MS stars out to
about 125~Kpc and to BSGs out to 500~Kpc. Only a few sources are seen in the
CMDs of the newly observed (C)HVCs with colors and magnitudes that would
correspond to such populations of stars. Therefore, we conclude that any
luminous young stellar populations contribute less than a few percent of the
stars to the near-infrared CMDs. 2MASS data are uniquely sensitive to Carbon
rich thermally pulsing AGB stars out to about 300~Kpc. Again, only a few
qualifying sources are seen in the CMDs of the (C)HVCs. The dozen sources that
could have been interpreted as AGB stars in two (C)HVCs were explained with
background clusters of galaxies above. Therefore, any luminous AGB stellar
population presumed to be intrinsic to the clouds contributes less than a few
percent of the stars on the CMDs of the fields.

While the data in principle allow for the presence of a few stars instrinsic
to each of the five (C)HVCs we observed, the implied characteristics of the
stellar content would be extremley different from that of all other dwarf
galaxies of the Local Group.  More likely, the (C)HVCs simply do not contain
any stars. We interpret our observations as non detections of stellar content
associated with the (C)HVCs.

\subsection{CHVC 224.0-83.4-197}

CHVC 224.0-83.4-197 deserves a few additional comments. It is located
on the sightline to the background Quasar Ton~S210, and has a
sufficient HI column density to produce Lyman series absorption lines
in the far ultraviolet. Sembach et al. (2002) recorded the absorption
spectrum of the CHVC, and used it to derive an upper limit to its
metallicity. As the upper limit is below that of the O/H ratio of the
local interstellar medium, they hypothesize that CHVC 224.0-83.4-197
could represent an extragalactic cloud of gas.

Our VLT observations do not include the sightline to the QSO, as they
were centered on the HI column density maximum of CHVC
224.0-83.4-197. The CMD of the FORS field is shown as Fig.~1. If the
clumping of faint stars near I of about 24.2 were interpreted as the
TRGB of a metal poor CHVC stellar population, then this would place
the cloud at a distance of more than 4~Mpc. We consider this to be an
unlikely explanation of the data, since the Galactic foregound
naturally provides an increase of sources toward faint magnitudes in
all of the CMDs observed.

The 2MASS source positions for the field surrounding the HI
maximum of CHVC 224.0-83.4-197 are shown in Fig.~2 and the
corresponding J-K$_S$ CMD is Fig.~3. The 2MASS field includes the
sightline to Ton~S210. There is nothing special associated in the
positional distribution of the sources near the QSO sightline. The
near-infrared CMD is not sensitive to a population of red giants at
Mpc distances and cannot be used to confirm that the faint optical
sources are due to stars at such a distance. However, just like the
optical CMD, the near-infrared CMD of CHVC 224.0-83.4-197 is naturally
explained with stars on a Galactic sightline at high latitudes, and
does not require the assumption of an extragalactic stellar content
associated with the cloud.

\section{Discussion}

We begin by discussing in how much the hypothesis that HVCs form the
repository of missing Local Group dwarf galaxies predicted by
cosmological simulations is constrained by the data. We have observed
ten (C)HVCs to similar limiting magnitudes. All ten cases are
interpreted as non detections of stellar content. We use binomial
statistic to ask what is the probability to get zero success in ten
trials. Specifically, the zero detection of stellar content in our
sample of ten HVCs can be used to set limits on the true fraction of
HVCs which may contain stars.  There is a 50\% chance of getting a
null result in ten trials if fewer than 7\% of all HVCs contains
stars. Assuming that the total sample of HVCs that can be interpreted
as infalling building blocks following Blitz et al. (1999) amounts to
582 (Putman 2006), we therefore cannot exclude the possibility that 41
HVCs are undiscovered dwarf galaxies of the Local Group.

The CMD approach to search for stars in HVCs has also been used by
other teams. Table~2 gives a list of (C)HVCs which have been observed
with single-star photometry in two or more filters, including
reference to the work, the limiting V or K magnitudes achieved by the
data, and the names of the clouds studied. All searches yielded null
results. The data are quite different in terms of limiting magnitudes
and sky coverage. The data for HVC~267+26+216 are too shallow to be
included in the statistical test. We also exlude the case of
Complex~H, which was observed in the near-infrared, only. This leaves
16 (C)HVCs with which we can test the Blitz hypothesis out to a
distance of about 600~Kpc. The non detections of stellar content in 16
(C)HVCs is consistent with the hypothesis that less than 4\% of HVCs
contain stars. Assuming that the 582 HVCs are uniformly distributed
throughout a volume with a 1~Mpc radius, the data cannot exlude the
possibility that 24 HVCs contain stars.

The observational biases for detecting dwarf galaxies around the Milky
Way in the optical were recently discussed by Willman et
al. (2004). The increased extinction and increase in Galactic
foreground stars toward the disk, combined with the magnitude limits
of surveys for overdensities of resolved stars, are predicted to
account for a total of 14-18 undetected dwarf satellites. Like our own
work, this estimate assumes that the properties of the missed
satellites are similar to those of the already known satellites. 

In summary, the observational selection effects indicate that a few
tens of dwarf galaxies may have been overlooked, but they cannot
account for a few hundred ``missing" dwarfs.

We now turn to the issue of the nature of HVCs. Unless the associated
stellar content is unlike that of any other known dwarf galaxy, we
must conclude that HVCs simply do not contain stars and are best
explained as pure gas clouds. Lacking the detection of stellar
``standard candles", the CMD work has not aided in determining the
distances of HVCs. However, recent indirect distance constraints for
HVCs have come from deep HI observations of M31 (Westmeier, Br\"uns,
and Kerp 2005) and of other galaxy groups (Zwann 2001, Pisano et
al. 2006). These works place the HVCs within the extended Galactic
halo, rather than throughout the Local Group. The ionization patterns
observed in a few highly ionized HVCs also suggest a location of at
least these clouds, in the Galactic halo or corona (e.g., Ganguly et
al. 2005).

\section{Conclusions}

The zero detection of stellar content in our sample of ten HVCs
tightens contraints on the true fraction of HVCs which may contain
stars. There is a 50\% chance of getting a null result in ten trials
if fewer than 7\% of all HVCs contains stars. We find that HVCs are a
very unlikely source to provide the ``missing" dwarf galaxies of the
Local Group which are predicted by cosmological simulations. It is
becoming very apparent that the ``missing" satellites have not simply
been overlooked observationally, indicating that observational bias is
not the easy way out of the substructure problem. If the predicted
minihalos do indeed exist, they do not contain stars. 

\section*{Acknowledgments}

We thank the ESO/VLT team for obtaining excellent service observations
for our project. This publication makes use of data products from the
Two Micron All Sky Survey. RS-L acknowledges funding through a NASA
Astrophysics Data Analysis Program grant awarded in response to the
2004 ROSS solicitation, and thanks the Max--Planck--Institut f\"ur
Extraterrestrische Physik in Garching for hosting a visit enabling our
team to complete this paper.

\bsp

\label{lastpage}


\begin{thebibliography}{99}

\bibitem{} Blitz, L., Spergel, D.N., Teuben, P.J., Hartmann, D.,
  Burton, W.B. 1999, ApJ, 514, 818 

\bibitem{} Braun, R. 2001, in ``Gas and Galaxy Evolution". ASP
  Conf. Proc. Vol. 240, p.479 (cf. also astro-ph/0009011)

\bibitem{} Braun, R., Burton, W.R. 1999, A\&A, 341, 437
 
\bibitem{} Davies, J., Sabatini, S., Davies, L., Linder, S., Roberts,
  S., Smith, R., Evans, Rh. 2002, MNRAS, 336, 155 

\bibitem{} Ganguly, R., Sembach, K. R., Tripp, T. M., Savage,
  B. D. 2005, ApJS, 157, 251 

\bibitem{} Hopp, U., Schulte-Ladbeck, R.E., Kerp, J. 2003, MNRAS, 399, 33
 
\bibitem{} Klypin, A., Kravtsov, A.V., Valenzuela, O., Prada, F. 1999,
  ApJ, 522, 82 

\bibitem{}Kuijken, K., Gilmore, G., 1989, MNRAS, 239, 571

\bibitem{} Lee, M.G., Freedman, W.L., Madore, B.F., 1993, ApJ, 417, 553

\bibitem{} Maraston, C. 1998, MNRAS, 300, 872

\bibitem{} Moore, B., Ghigna, S., Governato, F., Lake, G. et al. 1999,
  ApJL, 524, 19 

\bibitem{} Pisan, D.J., Barnes, D.G., Gibson, B.K., Staveley-Smith,
  L., Freeman, K.C., Kilborn, V.A. 2005, astro-ph/0509226 

\bibitem{} Putman, M.E., de Heij, V., Stavely-Smith, L., Braun, R.,
  Freeman, K.C., Gibson, B.K., Burton, W.B., Barnes, D.G., et
  al. 2002, AJ, 123, 873 

\bibitem{} Putman, M.E. 2006, Apj, in press (astro-ph/0603650)

\bibitem{} Robin, A.C., Reyle, C., Derriere, S., Picard, S., 2003,
  A\&A, 409, 523 

\bibitem{} Schlegel, D.J., Finkenbeiner, D.P., Davis, M. 1998, ApJ, 500, 525

\bibitem{} Sembach, K.R., Gibson, B.K., Fenner, Y., Putman, M.E. 2002,
  ApJ, 572, 178 

\bibitem{} Simon, J.D., Blitz, L. 2002, ApJ, 574, 726

\bibitem{} Simon, J.D., Blitz, L., Cole, A. A., Weinberg, M. D.,
  Cohen, M. 2006, ApJ, 640, 270 

\bibitem{} Siegel, M.H., Majewski, S.R., Gallart, C., Sohn, S.,
  Kunkel, W.E., Braun, R. 2005, ApJ, 623, 181 

\bibitem{} Wakker, B.P., van Woerden, H. 1997, ARA\&A, 35, 217

\bibitem{} Westmeier, T. 2003, Diploma thesis, Rheinische
  Friedrich-Wilhelms-Universit\"at Bonn 

\bibitem{} Westmeier, T., Br\"uns, C., Kerp, J. 2005, ASPC, 331, 105

\bibitem{} Willman, B., Dalcanton, J., Ivezic, Z., Schneider, D.P.,
  York, D.G. 2002, AJ, 124, 2600

\bibitem{} Willman, B., Governato, F., Dalcanton, J. J., Reed, D., Quinn, T. 2004, MNRAS, 353, 639

\bibitem{} Zwaan, M. A. 2001, MNRAS, 325, 1142

\end{thebibliography}
\end{document}